\documentstyle[12pt]{article}
\newcommand{\be}{\begin{equation}}
\newcommand{\ee}{\end{equation}}
\newcommand{\bea}{\begin{eqnarray}}
\newcommand{\eea}{\end{eqnarray}}
\newcommand{\ba}{\begin{array}}
\newcommand{\ea}{\end{array}}
\newcommand{\bmat}{\left(\ba}
\newcommand{\emat}{\ea\right)}

\newcommand{\pole}{\frac{1}{\varepsilon}\frac{g^2}{16\pi^2}}

\newcommand{\norsl}{\normalsize\sl}
\newcommand{\norsc}{\normalsize\sc}
\begin{document}
%-------------------- Title page ----------------------------------
\begin{titlepage}

\title{Renormalization of Twist-4 Operators in \\
       QCD Bjorken and Ellis-Jaffe Sum Rules
}

\author{
\norsc  Hiroyuki KAWAMURA\thanks{JSPS Research Fellow }  
\hspace{0.5cm} Tsuneo UEMATSU\thanks{Supported in part by
          the Monbusho Grant-in-Aid for Scientific Research
          No. C-06640392 } \\
\norsc  and Yoshiaki YASUI \\  
\norsl  Dept. of Fundamental Sciences\\
\norsl  FIHS, Kyoto University\\
\norsl  Kyoto 606-01, JAPAN\\
\\
\norsc  Jiro KODAIRA \\
\norsl  Dept. of Physics, Hiroshima University\\
\norsl  Higashi-Hiroshima 739, JAPAN\\
\\
}

\date{}

\maketitle

\begin{abstract}
{\normalsize
The QCD effects of twist-4 operators on the first moment of 
nucleon spin-dependent structure function $g_1(x, Q^2 )$ are studied 
in the framework of operator product expansion and renormalization group 
method. We investigate the operator mixing through renormalization
of the twist-4 operators including those proportional to the equation of 
motion by evaluating off-shell Green's functions 
%%%%%%%%%%%%%%%%%%%%%%%%added%%%%%%%%%%%%%%%%%%
in the usual covariant gauge as well as in the background gauge.
%%%%%%%%%%%%%%%%%%%%%%%%added%%%%%%%%%%%%%%%%%%
Through this procedure we extract the one-loop anomalous dimension of the 
spin 1 and twist-4 operator which determines the logarithmic correction to the
$1/Q^2$ behavior of the contribution from the twist-4 operators to the first
moment of $g_1(x,Q^2)$.}
\end{abstract}

\begin{picture}(5,2)(-310,-615)
\put(15,-100){KUCP-88-REV}
\put(15,-115){HUPD-9601-REV}
\put(15,-130){November 1996}
%\put(15,-135){revised}
\end{picture}

\vspace{2cm}
%\leftline{\hspace{1cm}hep-ph/9603338}
 
\thispagestyle{empty}
\end{titlepage}
\setcounter{page}{1}
\baselineskip 24pt

%----------------------- Text -----------------------------------

In the last several years there has been much interest in nucleon's spin 
structure functions $g_1(x,Q^2)$ and $g_2(x,Q^2)$, which can be measured by 
deep inelastic scattering of polarized leptons on polarized targets. 
Recent experiments on the nucleon spin structure functions carried out at 
CERN \cite{EMC,SMC} and SLAC \cite{E142,E143}, have stimulated intensive 
theoretical studies on the nucleon spin structure functions \cite{REV}.

In the deep inelastic scattering, the perturbative QCD has been tested so far
for the effects of the leading twist operators, namely twist-2 operators, for 
which the QCD parton picture holds.  Now the spin structure functions
would provide us with a good place to investigate higher-twist effects.
Our purpose in this paper is to study the renormalization of higher-twist
operators, especially the twist-4 operators, which are relevant for
the first moment of $g_1(x,Q^2)$, that corresponds to the Bjorken and
Ellis-Jaffe sum rules \cite{Bj,EJ}.
The anomalous dimension of the twist-4 operators determines the logarithmic 
correction to the $1/Q^2$ behavior of the twist-4 operator's contribution to 
the first moment of $g_1$.

The first moment of the $g_1(x,Q^2)$ structure functions for proton and
neutron turns out to be up to the power correction of order $1/Q^2$:
\bea
&&\hspace{-0.8cm}\Gamma_1^{p,n}(Q^2)\equiv \int_0^1 g_1^{p,n}(x,Q^2)dx 
\nonumber\\
&&\hspace{-0.8cm}=(\pm\frac{1}{12} g_A+\frac{1}{36}a_8)
\bigl ( 1-\frac{\alpha_s}{\pi}+ {\cal O}(\alpha_s^2)\bigr )
+\frac{1}{9}\Delta\Sigma\ \bigl ( 
1-\frac{33-8N_f}{33-2N_f}\frac{\alpha_s}{\pi}+{\cal O}(\alpha_s^2)
\bigr ),       
\eea
where $g_1^{p(n)}(x,Q^2)$ is the spin structure function of the 
proton (neutron) and the plus (minus) sign is for proton (neutron).
On the right-hand side, $g_A \equiv G_A/G_V$ is the ratio of the axial-vector 
to vector coupling constants. Here we assume that the number of active 
flavors in the current $Q^2$ region is $N_f=3$. Denoting
$\langle p,s|\overline{\psi}\gamma_\mu\gamma_5\psi|p,s\rangle
=\Delta q s_\mu$, the flavor-$SU(3)$ octet and singlet part, $a_8$ and 
$a_0=\Delta\Sigma$ are given by
\be
a_8 \equiv \Delta u +\Delta d -2\Delta s, \qquad
\Delta\Sigma \equiv \Delta u +\Delta d +\Delta s, \nonumber 
\ee
and $\Delta\Sigma$ is related to the scale-dependent density 
$\Delta\Sigma(Q^2)$ which evolves as
\be
\Delta\Sigma(Q^2)=\Delta\Sigma\left(1+\frac{6N_f}{33-2N_f}
\frac{\alpha_s(Q^2)}{\pi}\right),
\ee
hence $\Delta\Sigma$ is the density at $Q^2=\infty$.
Here we have suppressed the target mass effects, which can
be taken into account by the Nachtmann moments \cite{HKU}.
Note that taking the difference between $\Gamma_1^p$ and $\Gamma_1^n$ leads 
to the QCD Bjorken sum rule, the first order QCD correction of which 
was calculated in \cite{KMMSU,KMSU,K} and the higher order corrections were 
given in \cite{GL,LV,LTV,KS}. 

Now, the twist-4 operator gives rise to $O(1/Q^2)$ corrections \cite{QCD,DIQ}
to the first moment of $g_1(x,Q^2)$.  
As can be seen from the dimensional counting,  there is no contribution from 
the four-fermi type twist-4 operators to the first moment of $g_1(x,Q^2)$.
The only relevant twist-4 operators are of the form bilinear in 
quark fields and linear in the gluon field strength.
This is in contrast to the unpolarized case \cite{POLI}, where both types
of twist-4 operators contribute.
The common feature for the renormalization of higher-twist operator is that 
there appear a set of operators proportional to 
equations of motion, which we call EOM operators \cite{COLL,JL}. 
And there exists the operator mixing among twist-4 operators 
which can be studied in the off-shell Green's functions where the EOM 
operators are inevitable.
%including EOM 
%operators through renormalization. 
It should be emphasized that we have to keep the EOM operators 
to extract the physical observables like anomalous dimensions, 
which will be discussed later. 
%%%%%%%%%%%%%%%%%%%%%%%%%%%%%%%%%%%%%%%%%%%%%%%%%%%%%%%%%%%%%%%%%%%%%%%%%%%

The relevant operators in our case has the following properties;
The dimension of the operators is 5 and the spin is 1. Its parity is odd and 
it has to satisfy the charge conjugation invariance. The flavor non-singlet 
operators are bilinear in fermion fields. Here we have to consider gauge 
variant EOM operators as well.

Thus we have the following 
%five 
six operators which satisfy the above conditions:
\bea
&&  R_1^\sigma = -\overline{\psi}\gamma_5
       \gamma^{\sigma}D^2\psi, \quad                
  R_2^\sigma = g \overline{\psi}\tilde{G}^{\sigma\mu}\gamma_{\mu}\psi, 
           \nonumber\\
&&  E_1^\sigma = \overline{\psi} \gamma_5
          \not{\!\!D} \gamma^{\sigma} \not{\!\!D} \psi 
         -\overline{\psi} \gamma_5 D_\sigma\not{\!\!D}\psi
         -\overline{\psi} \gamma_5 \not{\!\!D} D^\sigma\psi,  \nonumber\\
&&  E_2^\sigma = \overline{\psi} \gamma_5
          \not{\!\!D} \gamma^{\sigma} \not{\!\!D} \psi 
         +\overline{\psi} \gamma_5 D_\sigma\not{\!\!D}\psi
         +\overline{\psi} \gamma_5 \not{\!\!D} D^\sigma\psi,\label{base}\\ 
&&  E_3^\sigma = \overline{\psi} \gamma_5\partial^{\sigma}
          \not{\!\!D}\psi + \overline{\psi} \gamma_5 \not{\!\!D}
          \partial^{\sigma}\psi,                          
\hspace{0.5cm}  E_4^\sigma = \overline{\psi} \gamma_5 \gamma^\sigma 
\not{\!\partial} \not{\!\!D}\psi + \overline{\psi} \gamma_5
\not{\!\!D}\not{\!\partial}\gamma^\sigma\psi,\nonumber
\eea
where $D_\mu=\partial_\mu-igA_\mu^aT^a$ is the covariant derivative and
$\tilde{G}_{\mu\nu}=\frac{1}{2}\varepsilon_{\mu\nu\alpha\beta}G^{\alpha\beta}$
is the dual field strength. And we work with massless quarks for simplicity 
of the argument.

Here one should note that not all of the above operators are independent, as 
in the case of twist-3 operators in $g_2(x,Q^2)$ \cite{KYU}, and they are
subject to the following constraint:
\be
R_1^{\sigma}=R_2^{\sigma}+E_1^{\sigma},
\ee
where we have used the identities, 
$D_\mu=\frac{1}{2}\{\gamma_\mu,\not{\!\!D}\}$ and
$[ D_\mu ,D_\nu ]=-igG_{\mu\nu}$.
Therefore any 
%five 
five operators out of (\ref{base}) are independent and
they mix through renormalization.

Here we take $(R_2,E_1,E_2,E_3,E_4)$ to be the basis of the independent 
operators.
The only operator which actually contributes to the physical matrix element 
responsible for the Bjorken sum rule is $R_2$.	
This twist-4 operator corresponds to the trace-part of twist-3 operator,
$(R_{\tau=3})_{\sigma\mu_1\mu_2}=g\overline{\psi}\tilde{G}_{\sigma\{\mu_1}
\gamma_{\mu_2\}}\psi-\mbox{traces}$,
but there is no relation between the basis for the twist-4 and that
for the twist-3 operators.

We now study the renormalization of the operators.
The composite operators, $O_i$, are renormalized by introducing the 
renormalization constants $Z_{ij}$ as
\be
(O_i)_R=\sum_j Z_{ij}(O_j)_B,
\ee
where the suffix $R$ ($B$) denotes renormalized (bare) quantities.
For the present basis we have the following renormalization mixing matrix:
\be
\bmat{c} 
R_2 \\ E_1 \\ E_2 \\ E_3 \\ E_4 
\emat_{\hspace{-0.1cm}R}
=
\bmat{ccccc}
Z_{11} & Z_{12} & Z_{13} & Z_{14} & Z_{15} \\
 0     & Z_{22} & Z_{23} & Z_{24} & Z_{25} \\
 0     & Z_{32} & Z_{33} & Z_{34} & Z_{35} \\
 0     & 0      & 0      &Z_{44}  & 0 \\
 0     & 0      & 0      &0       & Z_{55} 
\emat
\bmat{c} 
R_2 \\ E_1 \\ E_2 \\ E_3 \\ E_4
\emat_{\hspace{-0.1cm}B}.
\label{n3z1}
\ee

The general features for the mixing matrix are the following \cite{COLL,JL,H}:
(1)\ The counter terms for the EOM operators are supplied by the EOM 
operators themselves. This is because the on-shell matrix elements of 
the EOM operators ought to vanish.
(2)\ A certain type of operators do not get renormalized.
And if we take those operators as one of the independent base,
the calculation becomes much simpler.
(3)\ The gauge variant operators also contribute to the mixing.

We compute $Z_{ij}$ by evaluating
the off-shell Green's function of twist-4 composite operators
keeping the EOM operators as independent operators.
Thus we can avoid the subtle infrared divergence which may appear in the
on-shell amplitude with massless particle in the external lines.
Another advantage to study the off-shell Green's function is that
we can keep the information on the operator mixing problem.
And further, the calculation is much more straightforward than the one 
using the on-shell conditions.

At the tree level, $R_2$ operator 
%does not contribute to the 2-point functions, but it 
contributes to the 3-point functions with quarks $\psi$, 
$\overline{\psi}$ and a gluon, $A_\mu$ in the external lines. 
So we consider the following one-particle irreducible (1PI) 
Green's function:
\bea
\Gamma_{O_\sigma}^{\psi\bar{\psi}A}\equiv \langle 0 |
T(O_\sigma\psi(p')A_{\rho}^a(l)\bar{\psi}(p))|0 \rangle^{1PI} ,
\label{eight}
\eea
where the fields and the coupling constant involved represent the bare 
quantities. Here we employ the dimensional regularization 
($D=4-2\varepsilon$) and take the minimal subtraction scheme. 
The Green's functions are renormalized as follows:
\be
{\left(\Gamma_{O_i}\right)}_R =\sum_jZ_2\sqrt{Z_3}Z_{ij}
(\Gamma_{O_j})_B,
\ee
where $Z_2$ and $Z_3$ are wave function renormalization constants for
quarks and gluon fields.
%In Feynman gauge, the renormalized and bare coupling constants are related as 
%\be
%Z_2Z_3^{1/2}\gb=\left\{1-\pole [C_2(R)+C_2(G)]\right\}\gr,
%\ee
%where the quadratic Casimir operators are $C_2(R)=4/3$ and $C_2(G)=3$ for QCD.
%%%%%%%%%%%%%%%%%%%%%%%%%%%%%added%%%%%%%%%%%%%%%%%%%%%%%%%%%%%%%%%%%
We first present the evaluation in the usual covariant gauge. 
%%%%%%%%%%%%%%%%%%%%%%%%%%%%%added%%%%%%%%%%%%%%%%%%%%%%%%%%%%%%%%%%%
%Combined with the conditions from the 2-point functions, 
The one-loop radiative corrections arising from eight diagrams for $R_2$
are represented as:
\bea
(\Gamma_{R_2}^{\psi\bar{\psi}A})_{\mbox{1-loop}}&=&
\left\{ 1+\pole [-\frac 5 3 C_2(R)+C_2(G)]\right\} 
(\Gamma_{R_2}^{\psi\bar{\psi}A})_{\mbox{tree}} \nonumber\\
&+& \pole [-\frac 3 2 C_2(R)+\frac 3 8  C_2(G)]
(\Gamma_{E_1}^{\psi\bar{\psi}A})_{\mbox{tree}} \nonumber\\
&+& \pole [-\frac 1 6  C_2(R)+\frac 1 8 C_2(G)]
(\Gamma_{E_2}^{\psi\bar{\psi}A})_{\mbox{tree}} \nonumber\\
&+& \pole[- \frac 1 4 C_2(G)](\Gamma_{E_3}^{\psi\bar{\psi}A})_{\mbox{tree}}\label{GR2}\\
&+& \pole \frac 1 4 C_2(G)(\Gamma_{E_4}^{\psi\bar{\psi}A})_{\mbox{tree}},\nonumber
\eea
where the quadratic Casimir operators are $C_2(R)=4/3$ and $C_2(G)=3$ for QCD.
%%%%%%%%%%%%%%%%%%%%%%% added paragraph %%%%%%%%%%%%%%%%%%%%%%%%%%%%%%%%
In (\ref{GR2}), the tree-level Green's functions are given by
\bea
(\Gamma_{R_2}^{\psi\bar{\psi}A})_{\mbox{tree}}&=&
ig\varepsilon_{\sigma\rho\alpha\beta}l^{\alpha}\gamma^{\beta}T^a,
\nonumber\\
(\Gamma_{E_1}^{\psi\bar{\psi}A})_{\mbox{tree}}&=&
g\gamma_5\gamma_{\sigma}(p+p')_{\rho}T^a-
ig\varepsilon_{\sigma\rho\alpha\beta}l^{\alpha}\gamma^{\beta}T^a,
\nonumber\\
(\Gamma_{E_2}^{\psi\bar{\psi}A})_{\mbox{tree}}&=&
-2g\gamma_5g_{\sigma\rho}(\not{\! p}+\not{\! p}')T^a
-2g\gamma_5\gamma_{\rho}(p+p')_{\sigma}T^a \nonumber\\
&&+g\gamma_5\gamma_{\sigma}(p+p')_{\rho}T^a-
ig\varepsilon_{\sigma\rho\alpha\beta}l^{\alpha}\gamma^{\beta}T^a\nonumber\\
(\Gamma_{E_3}^{\psi\bar{\psi}A})_{\mbox{tree}}&=&
-g\gamma_5\gamma_{\rho}(p+p')_{\sigma}T^a,
\nonumber\\
(\Gamma_{E_4}^{\psi\bar{\psi}A})_{\mbox{tree}}&=&
g\gamma_5g_{\sigma\rho}(\not{\! p}+\not{\! p}')T^a
-g\gamma_5\gamma_{\rho}(p+p')_{\sigma}T^a \nonumber\\
&&-g\gamma_5\gamma_{\sigma}(p+p')_{\rho}T^a+
ig\varepsilon_{\sigma\rho\alpha\beta}l^{\alpha}\gamma^{\beta}T^a
\eea
%%%%%%%%%%%%%%%%%%%% end of added paragraph %%%%%%%%%%%%%%%%%%%%%%%%%%%%
%%%%%%%%%%%%%%%%begin added%%%%%%%%%%%%%%%%%%%%%%%%%%%%%%%%%%%%%%%%%%%%%%%%
Here one can easily see that these five operators have their 
tree-level 3-point functions as linear combinations 
of four independent tensor structures. 
So in order to identify the counter terms properly as given in (\ref{GR2}) 
we need to make use of the conditions for $Z_{ij}$ extracted 
from the 2-point functions with $\psi$, $\overline{\psi}$ in the external 
lines.
%%%%%%%%%%%%%%%%end added%%%%%%%%%%%%%%%%%%%%%%%%%%%%%

Note that the tree-level tensor structure for $R_2$, 
$ig\varepsilon_{\sigma\rho\alpha\beta}l^{\alpha}\gamma^{\beta}T^a$,
appears also in those for $E_1$, $E_2$ and $E_4$. 
Therefore, in order to extract 
the correct mixing-matrix element, it is crucial to keep the EOM operators.
This feature is quite in contrast to the case of twist-2  operators, where
we do not have to consider EOM operators at all. 

%Writing the renormalization constants as
%\bea
%Z_{ij}\equiv \delta_{ij} +
%{1\over\varepsilon}{{g^2}\over{16\pi^2}}z_{ij},
%\eea
%we have
%\bea
%z_{11}&=&\frac 8 3 C_2(R), \quad z_{12}=\frac 4 3 C_2(R)-\frac 1 4 C_2(R),
%\nonumber\\
%z_{13}&=&\frac 2 3 C_2(R)-\frac 1 4 C_2(G), \quad z_{14}=-\frac 1 3 C_2(R).
%\eea

For the Green's functions of the EOM operators, we have additional Feynman 
diagrams due to the presence of the two-point vertices at the tree level.
Further, the EOM operators like $E_3$ and $E_4$ which are of the form 
$E=\overline{\psi}B\displaystyle{\frac{\delta S}{\delta\overline{\psi}}}$, 
where $B$ is independent of fields,  do not get renormalized:
$Z_{44}=Z_{55}=1$.

To summarize we get the following result for the renormalization constants.
(The detailed calculation will be discussed elsewhere \cite{Kawa}):
\bea
\ba{ll}
z_{11}={8\over 3}C_2(R),  & z_{12}={3\over 2}C_2(R)-{3\over 8}C_2(G), \\ 
z_{13}={1\over 6}C_2(R)-{1\over 8}C_2(G), & z_{14}={1\over 4}C_2(G), \\
z_{15}=-{1\over 4}C_2(G), & z_{22}={1\over 2}C_2(R)+{3\over 8}C_2(G),\\
z_{23}=-{1\over 2}C_2(R)-{1\over 8}C_2(G),& z_{24}={1\over 4}C_2(G)\\ 
z_{25}={1\over 8}C_2(G),  & z_{32}=-{3\over 2}C_2(R)-{3\over 8}C_2(G),\\
z_{33}=-{1\over 2}C_2(R)+{1\over 8}C_2(G), & z_{34}=-{1\over 4}C_2(G), \\
z_{35}=-{1\over 8}C_2(G), & z_{44}=z_{55}=0 .
\label{rm}
\ea
\eea
where we have written the renormalization constants as
\be
Z_{ij}\equiv \delta_{ij} +
{1\over\varepsilon}{{g^2}\over{16\pi^2}}z_{ij}.
\ee
This result is in agreement with the general theorem on the renormalization
mixing matrix discussed above.

We now determine the anomalous dimension of $R_2^{\sigma}$ operator. 
In physical matrix elements, the EOM operators do not contribute \cite{POLI}
and we have
\be
\langle\mbox{phys}|(R_2^{\sigma})_B|\mbox{phys}\rangle =Z_{11}^{-1}
\langle\mbox{phys}|(R_2^{\sigma})_R|\mbox{phys}\rangle 
=( 1-\frac{g^2}{16\pi^2}\frac{1}{\epsilon}\frac{8}{3}C_2(R)) 
\langle\mbox{phys}|(R_2^{\sigma})_R|\mbox{phys}\rangle.
\ee
Therefore  the anomalous dimension $\gamma_{R_2}$ turns out to be
\be
\gamma_{R_2}(g)\equiv Z_{11}\mu\frac{d}{d\mu}(Z_{11}^{-1}) 
=\frac{g^2}{16\pi^2}\gamma_{R_2}^0+O(g^4), \quad
\gamma_{R_2}^0=2z_{11}=\frac{16}{3}C_2(R), \label{anomdim}
\ee
which coincides with the result obtained by Shuryak and Vainshtein \cite{SV}
%in a different method.
%%%%%%%%%%% begin of modified and added paragraph %%%%%%%%%%%%%%% 
based on the background field method \cite{SV1} in the coordinate space,
where they discarded the contribution from the EOM operators by taking the
on-shell quark external states using the equations of motion for massless
quarks given by
$$\not{\!\!D} \psi=\overline{\psi}\not{\!\!D}=0.$$
%, where the gluon field in the
%twist-4 operator, $R_2^{\sigma}$, is treated as classical background field,
%$A_\mu^{cl}$, in the leading order.  So they only consider the effective 
%action with two external quark lines. They work in the coordinate space, 
%where the equations of motion for massless quarks:
%$\not{\!\!D} \psi=\overline{\psi}\not{\!\!D}=0$, are most directly expressed.
%They could discard the contribution from the EOM operators by taking 
%quark external states on-shell using the above equations of motion. 
%%%%%%%%%%% end of modified and added paragraph %%%%%%%%%%%%%%%
%%%%%%%%%%%%%%%begin of further added paragraph%%%%%%%%%%%%%%%%%%%%%%%%
Here we also presents our result for the renomalization mixing of 
the twist-4 operators in the background field method \cite{Abbot}. 
We shall work with the momentum space. 
In this method we decompose the gauge field into classical background
field and the quantum field as:
$$
A_\mu^a=A_\mu^{a(cl)}+a_\mu^a,
$$
and set up the Feynman rule, where we have an additional term in the 
three-gluon vertex \cite{Abbot} contributing to this calculations.
In the background field method, there appear only gauge invariant
operators contributing the mixing through renormalization \cite{KZ}. 
We take the independent operator basis to the three gauge invariant operators;
$R_2$, $E_1$ and $E_2$. 
Here we calculated the Green's function (\ref{eight}) with $A_\mu^{cl}$
as the external gauge field.
Taking into account the wave function renormalization constant of the 
background gauge field, we obtain the renormalization mixing matrix:
\be
\bmat{c} 
R_2 \\ E_1 \\ E_2  
\emat_{\hspace{-0.1cm}R}
=
\bmat{ccc}
1+\frac{8}{3}C_2(R){\hat\alpha}/{\varepsilon}
 & \frac{3}{2}C_2(R){\hat\alpha}/{\varepsilon} 
& \frac{1}{6}C_2(R){\hat\alpha}/{\varepsilon} \\
 0     & 1+\frac{1}{2}C_2(R){\hat\alpha}/{\varepsilon} 
& -\frac{1}{2}C_2(R){\hat\alpha}/{\varepsilon} \\
 0     & -\frac{3}{2}C_2(R){\hat\alpha}/{\varepsilon} 
& 1-\frac{1}{2}C_2(R){\hat\alpha}/{\varepsilon} 
\emat
\bmat{c} 
R_2 \\ E_1 \\ E_2 
\emat_{\hspace{-0.1cm}B},
\label{bg}
\ee
where ${\hat\alpha}=g^2/16\pi^2$.
This result leads to the same physically observable anomalous dimension
of $R_2$ as given in (\ref{anomdim}).

%%%%%%%%%%%%%%%end of further added paragraph%%%%%%%%%%%%%%%%%%%%%%%%

Including the twist-4 effect the Bjorken sum rule becomes 
\bea
&&\int_0^1 dx\left[g_1^p(x,Q^2)-g_1^n(x,Q^2) \right] \nonumber\\
&=&
\frac{1}{6}\left\{g_A\left(1-\frac{\alpha_s(Q^2)}{\pi}+{\cal O}(\alpha_s^2)
\right) 
-\frac{8}{9Q^2}f_3
\left\{\frac{\alpha_s(Q_0^2)}{\alpha_s(Q^2)}
\right\}^{-\displaystyle{32/9\beta_0}}\right\},
\eea
where $f_3$ is the reduced matrix element of $R_{2\sigma}^3$,
renormalized at $Q_0^2$, which is defined for the general flavor
indices, with $t^i$ being the flavor matrices, as 
\be
R_{2\sigma}^i=g\overline{\psi}\tilde{G}_{\sigma\nu}\gamma^{\nu}t^i\psi,
\quad \langle p,s|R_{2\sigma}^i|p,s \rangle=f_i s_{\sigma} 
\quad (i=0,\cdots,8).
\ee

So far we have considered the flavor non-singlet part. Now we turn to the
flavor singlet component.
Here we note that there is only one non-vanishing independent gluon operator: 
$\tilde{G}^{\alpha\sigma}{D}^{\mu}G_{\mu\alpha}$. 
This operator is equal to the flavor-singlet operator 
$R_{2\sigma}^{\scriptscriptstyle{0}}$ up to  
the gluon's equation of motion:  
\be
\tilde{G}^{\alpha\sigma}{D}^{\mu}G_{\mu\alpha}=g\overline{\psi}\gamma_{\alpha}
\tilde{G}^{\sigma\alpha}\psi.
\ee
So now we have only to take into account the mixing between 
$R_{2\sigma}^{\scriptscriptstyle{0}}
=g\overline{\psi}\gamma_{\alpha}\tilde{G}^{\sigma\alpha}\psi$
and
\be
E_G^{\sigma}=\tilde{G}^{\alpha\sigma}D^{\mu}G_{\mu\alpha}-
g\overline{\psi}\gamma_{\alpha}\tilde{G}^{\sigma\alpha}\psi,
\ee
in addition to the previous results for the non-singlet part. 
The mixing between $R_2^{\scriptscriptstyle{0}}$ and $E_G$ can be studied 
by computing the Green's function with two-gluon external lines,
$\Gamma_{R_2^0}^{\ AA}$,
shown in Fig.1.  Now we introduce the renormalization constant $Z_{16}$ as 
\be
\left(R_2\right)_R=Z_{11}\left(R_2\right)_B+Z_{12}\left(E_1\right)_B
+Z_{13}\left(E_2\right)_B+Z_{14}\left(E_3\right)_B
+Z_{15}\left(E_4\right)_B+Z_{16}\left(E_G\right)_B.
\ee
From the diagrams of Fig.1, we get for the number of flavors $N_f$:
\be
Z_{16}=\pole\times\frac 2 3 N_f,
\ee
hence we obtain the exponent for the singlet part
\be
-\frac{\gamma_S^0}{2\beta_0}=-\frac{\gamma_{NS}^0}{2\beta_0}
-\frac{2}{3}\frac{N_f}{\beta_0}
=-\frac {1}{\beta_0}\left(\frac{32}{9} + \frac{2}{3}N_f\right).
\ee
Including the twist-4 effects the first moment of $g_1^{p,n}(x,Q^2)$ becomes
\bea
&&\hspace{-0.8cm}\Gamma_1^{p,n}(Q^2)\equiv \int_0^1 g_1^{p,n}(x,Q^2)dx 
\nonumber\\
&&\hspace{-0.8cm}=(\pm\frac{1}{12} g_A+\frac{1}{36}a_8)
\bigl ( 1-\frac{\alpha_s}{\pi}+ {\cal O}(\alpha_s^2)\bigr )
+\frac{1}{9}\Delta\Sigma\bigl ( 
1-\frac{33-8N_f}{33-2N_f}\frac{\alpha_s}{\pi}+{\cal O}(\alpha_s^2)
\bigr )  \nonumber\\     
&&\hspace{-0.8cm}-\frac{8}{9Q^2}\Bigl [ \{ \pm\frac{1}{12} f_3+
\frac{1}{36}f_8\}
\left(\frac{\alpha_s(Q_0^2)}{\alpha_s(Q^2)}\right)^
{-\frac{\gamma_{NS}^0}{2\beta_0}}\hspace{-0.1cm}+\frac{1}{9}f_0
\left(\frac{\alpha_s(Q_0^2)}{\alpha_s(Q^2)}\right)^
{-\frac{1}{2\beta_0}
(\gamma_{NS}^0+\frac{4}{3}N_f)}\Bigr ], \label{mom}
\eea
where $f_0$, $f_3$ and $f_8$ are the twist-4 counter parts of $a_0$, $a_3$ 
and $a_8$. $f_i$'s are scale dependent and here they are those at $Q_0^2$.

If we take into account the ghost terms in our QCD lagrangian, we get extra
terms for the gluon EOM operator, which are expressed 
in terms of the ghost fields and
satisfy the BRST invariance. In addition, there appears the so-called BRST 
exact operator \cite{COLL,JL,H} which participates in the operator mixing. 
However, it turns out that their contributions cancel with each other, and 
the final result does not change \cite{Kawa}. This can be more easily seen
in the background gauge where we have only $E_G$ for the additional 
independent operator and no ghost fields.

Finally it should be noted that the matrix elements of the twist-4 
operators $f_i$'s are 
considered to have ambiguities due to the renormalon singularity 
as discussed in the literatures \cite{Renom}.
However, the exponents of logarithmic corrections to the $1/Q^2$ behavior, 
which we computed in the present paper, have definite values without any
ambiguity. In case the $Q^2$ dependence of the moment (\ref{mom}) 
could be measured with enough accuracy in future experiments, we would be 
able to examine the presence of the twist-4 effects. 
 
\vspace{0.5cm}
We would like to thank K. Tanaka for valuable discussion.
%%%%%%%%%%%%%%%%%%%%%%%%%%%%%%%%%%%%%%%%%%%%%%

%%%%%%%%%%%%%%%%%%%% Figure caption %%%%%%%%%%%%%%%%%%
%\newpage
\vspace{3cm}
\noindent
{\large Figure Caption}
\baselineskip 16pt

\vspace{0.5cm}
\noindent
Fig.1 \quad 
The Feynman diagrams for $\Gamma_{R_2^0}^{\ AA}$ contributing to 
the mixing of $R_2^{\scriptscriptstyle{0}}$ and $E_G$.

\newpage

\input epsf.sty
\begin{figure}
\centerline{
\epsfxsize=11cm
\epsfbox{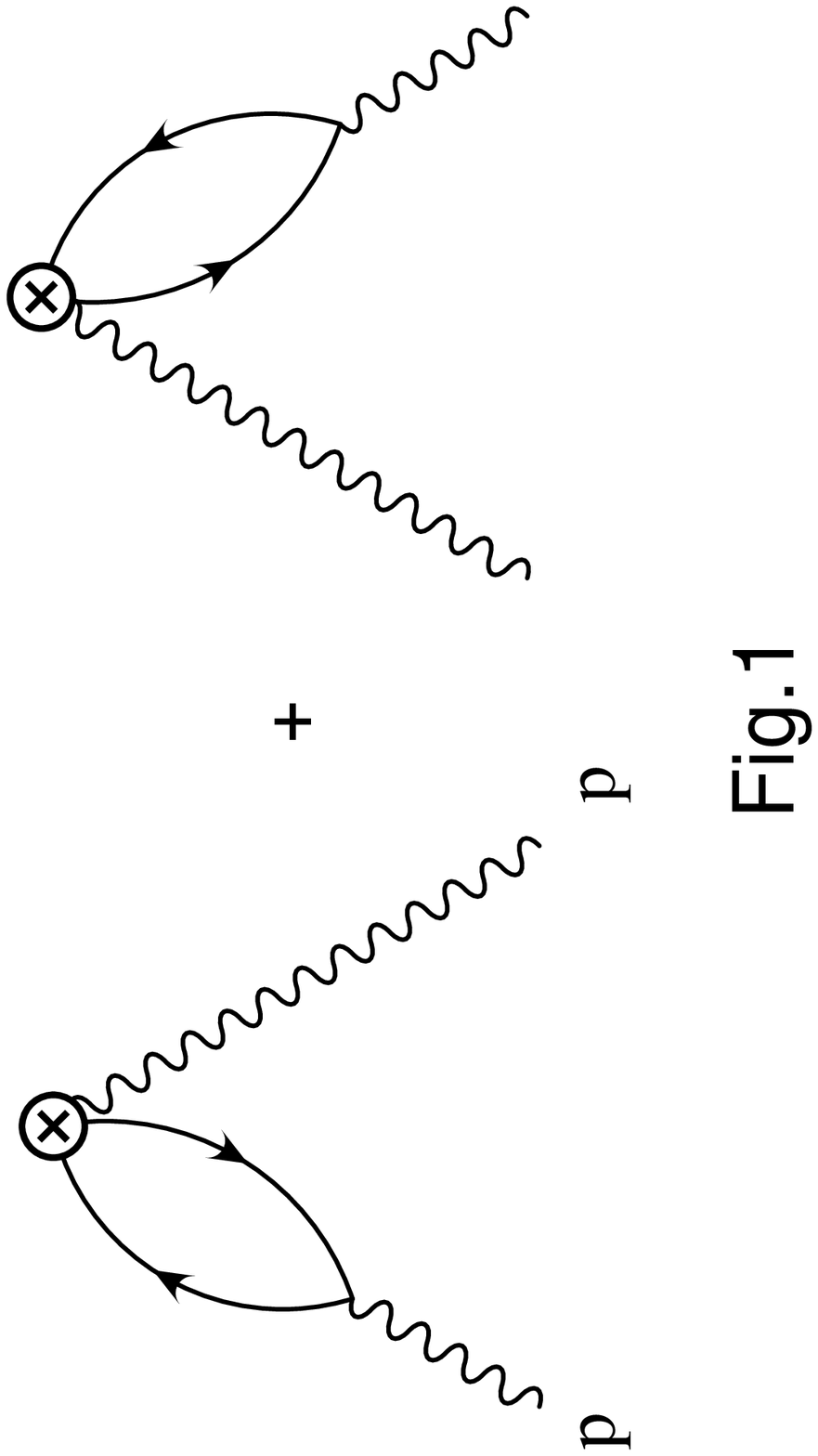}}
\end{figure}

\end{document}